\documentclass[iop]{emulateapj}

\usepackage{amsmath,natbib,graphicx}
\usepackage{epsf}
\usepackage{epstopdf}
\usepackage{epsfig}
\usepackage{color}
\DeclareGraphicsExtensions{.jpg,.pdf,.png,.eps,.ps}
\usepackage{scrextend}
\usepackage{hyperref}

\newcommand{\kms}{km $s^{-1}$}
\newcommand{\II}{~{\sc ii}}
\newcommand{\I}{~{\sc i}}
\newcommand{\III}{~{\sc iii}}

\begin{document}

\title{Revealing the host galaxy of a quasar 2175 \AA$ $ dust absorber at $z$ = 2.12}



\shortauthors{J. Ma, et al.}

\author{
Jingzhe Ma$^{1,2}$, Gabriel Brammer$^{3}$, Jian Ge$^{1}$, J. Xavier Prochaska$^{4}$, and Britt Lundgren$^{5}$}

\altaffiltext{1}{Department of Astronomy, University of Florida, Gainesville, FL 32611, USA}
\altaffiltext{2}{Department of Physics \& Astronomy, University of California, Irvine, CA 92617, USA; \href{mailto:jingzhem@uci.edu}{jingzhem@uci.edu}}
\altaffiltext{3}{Space Telescope Science Institute, 3700 San Martin Dr, Baltimore, MD 21211, USA}
\altaffiltext{4}{Department of Astronomy and Astrophysics, UCO/Lick Observatory, University of California, 1156 High Street, Santa Cruz, CL 95064, USA}
\altaffiltext{5}{Department of Physics, University of North Carolina, Asheville, NC 28804, USA}


\begin{abstract}

We report the first detection of the host galaxy of a strong 2175 \AA$ $ dust absorber at $z$ = 2.12 towards the background quasar SDSS J121143.42+083349.7 using {\it HST}/WFC3 IR F140W direct imaging and G141 grism spectroscopy. The spectroscopically confirmed host galaxy is located at a small impact parameter of $\sim$ 5.5 kpc ($\sim$ 0.65$''$). The F140W image reveals a disk-like morphology with an effective radius of 2.24 $\pm$ 0.08 kpc. The extracted 1D spectrum is dominated by a continuum with weak emission lines ([O\III] and [O\II]). The [O\III]-based unobscured star formation rate (SFR) is 9.4 $\pm$ 2.6  M$_{\odot}$yr$^{-1}$ assuming an [O\III]/H$\alpha$ ratio of 1. The moderate 4000 \AA$ $ break (Dn(4000) index $\sim$ 1.3) and Balmer absorption lines indicate that the host galaxy contains an evolved stellar population with an estimated stellar mass M$_*$ of (3 - 7) $\times$ 10$^{10}$ M$_{\odot}$. The SFR and M$_*$ of the host galaxy are comparable to,  though slightly lower than, those of typical emission-selected galaxies at $z$ $\sim$ 2. As inferred from our absorption analysis in \cite{Ma2015, Ma2017, Ma2018}, the host galaxy is confirmed to be a chemically-enriched, evolved, massive, and star-forming disk-like galaxy that is likely in the transition from a blue star-forming galaxy to a red quiescent galaxy. 

\end{abstract}


\section{Introduction}\label{sec:intro}

Absorption line systems towards bright background quasars or gamma-ray burst afterglows provide excellent probes of the gas in and around galaxies irrespective of their luminosities. The absorber-galaxy relation is established by detecting the emission counterparts of absorption systems. The searches for emission counterparts of the extensively studied damped Lyman-$\alpha$ absorbers (DLAs) and Mg\II{} absorbers have been conducted for more than two decades (see recent reviews \citealt{Krogager2017,Joshi2017}). The picture of what types of galaxies hosting these absorbers is emerging with an increased number of detections, although the detection rate remains very low largely due to the fact that the absorber's host tends to be too faint or too close in angular separation to the the bright quasar. DLA counterparts found in emission have properties overlapping with Lyman-$\alpha$ emitters (LAEs), Lyman break galaxies (LBGs), and dwarf galaxies (e.g., \citealt{Fynbo2008, Rauch2008, Krogager2013,Fumagalli2015, Neeleman2017}). A more general paradigm is that DLAs trace star-forming galaxies with a large span of masses, luminosities, and star formation rates (SFRs). Mg\II{} absorbers are found to be associated with either galactic-scale outflows originating from their host galaxies \citep{Bouche2006,Lundgren2012,Bordoloi2014}, dynamical mergers or filamentary accretion onto galaxies \citep{Steidel2002,Chen2010,Rubin2012}, or high-velocity clouds \citep{Richter2012,Herenz2013}. Therefore Mg\II{} absorbers can be used to trace many different feedback processes regulating the evolution of their host galaxies \citep{Joshi2017}.

A new population of quasar absorption systems beyond the local Universe, 2175 \AA$ $ dust absorbers (2DAs; the broad absorption feature centered around rest-frame 2175 \AA$ $ commonly seen in the Milky Way extinction curves; e.g. \citealt{Wang2004,Srianand2008, Jiang2011}), and its relationships with other absorption line systems (e.g., DLAs, Mg\II{} absorbers) has been established \citep{Ma2018}. These absorbers contain signatures of not only gas but also dust in the absorbing galaxies \citep{Ma2017,Ma2018}. 2DAs are a subset of Mg\II$ $ (and Fe\II) absorbers. Compared with DLAs or subDLAs in the literature, the 2DAs on average have higher metallicities, higher depletion levels (onto dust), and larger velocity widths, indicating higher stellar masses (with an estimated median value of $\sim$ 2 $\times$ 10$^{10}$ M$_{\odot}$; \citealt{Ma2018}). They are more likely to be drawn from the same parent population as metal-strong DLAs/subDLAs, which are defined as having logN(Zn\II/cm$^{-2}$) $\geq$ 13.15 or logN(Si\II/cm$^{-2}$) $\geq$ 15.95 \citep{Herbert-Fort2006}. In addition to a great number of low ionization lines (e.g., Zn\II, Mg\II, Fe\II, Si\II, Al\II, Ni\II, Mn\II, Ca\II, Ti\II) detected in 2DAs, neutral C\I$ $ lines are detected in all of the 2DAs (as long as covered in the spectra) with logN(C\I/cm$^{-2}$) $>$ 14.0 \citep{Ma2018}. Given the relatively high metallicity and the presence of C\I$ $, 2DAs can be used as molecular gas tracers, and the simultaneous presence of the 2175 \AA$ $ bump and molecular gas (H$_2$ and/or CO) has been reported in a few absorbers (e.g., \citealt{Noterdaeme2009, Prochaska2009, Ma2015, Noterdaeme2017}). 

All the pieces of evidence suggest that the host galaxies are likely to be chemically enriched, evolved, massive (more massive than typical DLA/subDLA galaxies), and presumably star-forming galaxies given the neutral and molecular gas in the systems. To establish the absorber-galaxy relation for 2DAs and obtain fundamental galaxy properties such as star formation rate (SFR) and stellar mass, we need to reveal the host galaxies in emission. The emission counterparts of 2DAs are expected to be detected more easily than typical DLAs/subDLAs \citep{Ma2018}. Confirming the emission counterparts of 2DAs is essential to placing this population in the context of galaxy populations in general. Upon establishing this association, we may leverage the exquisite measurements of these 2DAs to examine the chemically mature interstellar medium (ISM) of galaxies in the $z$ $\sim$ 2 universe.

The strong 2DA at $z$ = 2.1166 towards QSO J121143.42+083349.7 ($z$ = 2.4828, hereafter J1211+0833) from the Sloan Digital Sky Survey Data Release 10 stands out as an excellent target for revealing the host galaxy in emission. It is classified as a solar-metallicity DLA and a Mg\II$ $ (and Fe\II) absorber with the remarkably simultaneous presence of C\I, CO and neutral chlorine (Cl\I; a tracer of H$_2$). The chemical abundances and physical conditions are characterized in detail in \cite{Ma2015}. 

In this Letter, we report the detection of the host galaxy of this 2DA using {\it HST}/WFC3 IR direct imaging and slitless spectroscopy. Section \ref{sec:observations} and Section \ref{sec:reduction} describe the observations and data reduction processes that lead to the confirmation of the galaxy redshift. We discuss the properties derived from the {\it HST} data in Section \ref{sec:discussion} and summarize the results in Section \ref{sec:conclusion}. Throughout this paper, we assume a flat $\Lambda$CDM cosmological model with H$_0$ = 70 \kms Mpc$^{-1}$, $\Omega$$_{M}$ = 0.3, and $\Omega$$_{\Lambda}$ = 0.7.

\section{Observations and initial data reduction} \label{sec:observations}

The WFC3 G141 grism has spectral coverage from 1.1 to 1.65 $\micron$ with a dispersion of 46.5 \AA$ $ pixel$^{-1}$ ($R$ $\sim$ 130) in the primary spectral order before drizzling. Thus, the grism observations cover the rest-frame Balmer break, 4000 \AA$ $ break, 3727 \AA$ $ [O II], H$\beta$, and 4959 and 5007 \AA$ $ [O III] features at the redshift of the absorber's host galaxy.

\begin{figure}
\centering
{\includegraphics[width=7cm, height=6.9cm]{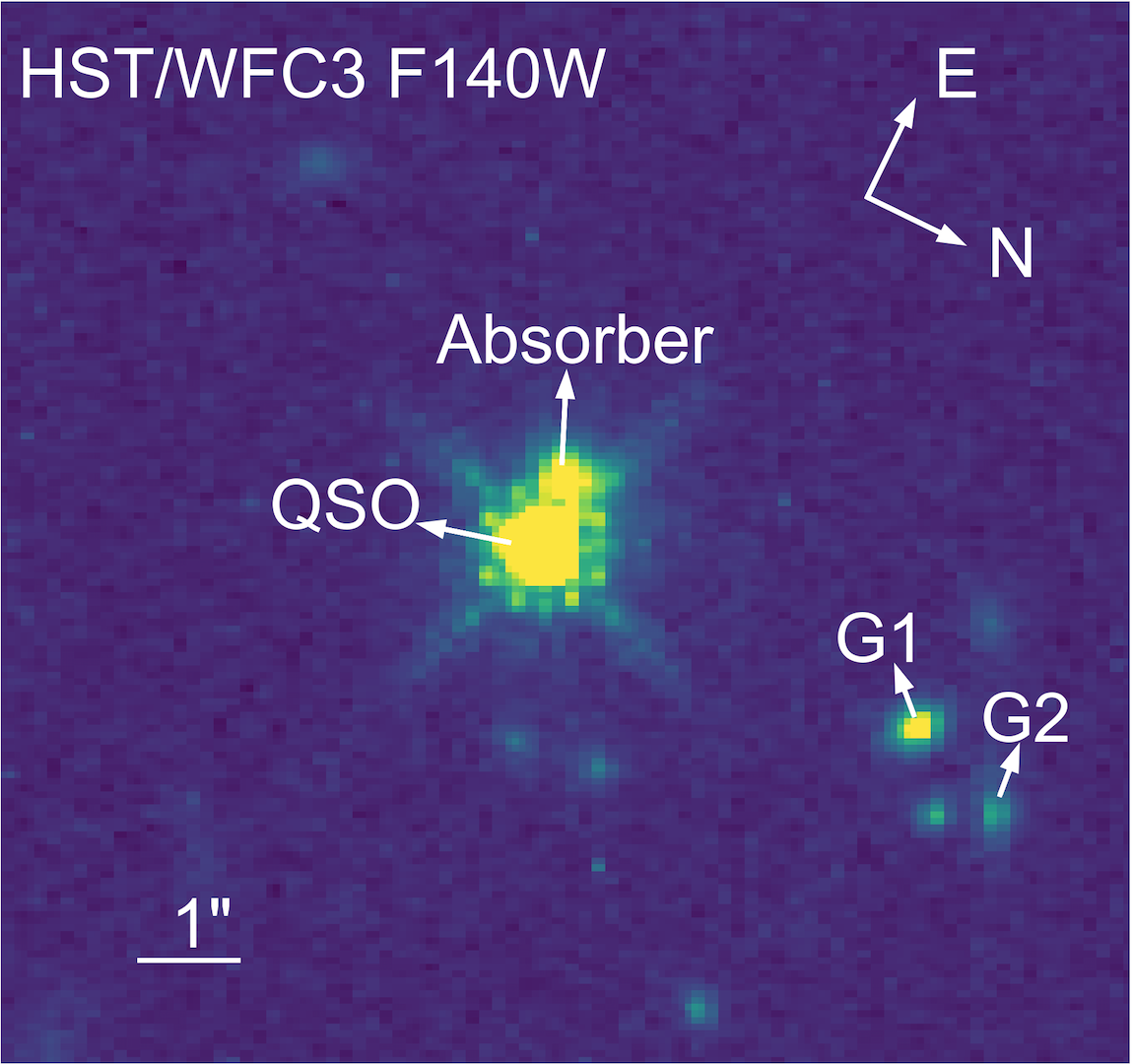}}   
\caption{HST/WFC3 F140W image. The host galaxy of the J1211+0833 absorber is revealed $\sim$ 0.65$''$ from the quasar sightline.  G1 and G2 are two nearby galaxies at $z$ = 2.1188 and $z$ = 2.2433, respectively.  }
\label{fig:F140W}
\end{figure}

 This target was observed with paired F140W direct imaging and G141 grism in three visits (one orbit per visit) on 15 November 2015, 24 December 2015, and 26 February 2016 with dispersion position angles, 75$^\circ$, 55$^\circ$, and 26$^\circ$ east of north, respectively (GO 14200, PI: J. Ma). Two dithered pairs of direct (F140W, 465 s) and grism (G141, 2206 s) images were obtained. The direct images are used for source identification (Figure \ref{fig:F140W}) and wavelength calibration of the slitless spectra.

The direct imaging and grism data were reduced together using the ``Grism redshift \& line" analysis software for space-based slitless spectroscopy (Grizli\footnote{https://github.com/gbrammer/grizli/}; Brammer et al. in preparation). The end-to-end processing of the data includes four major steps: (1) pre-processing of FLT files downloaded from MAST, (2) field contamination modeling, (3) spectral extractions, (4) redshift and continuum \& emission line fitting. The pre-processing involves astrometric registration, grism sky background subtraction \& flat-fielding, and extracting visit-level catalogs and segmentation images from the direct imaging \citep{Brammer2012}. The products of this step are aligned, background-subtracted FLTs and drizzled mosaics of direct and grism images (0.06$''$ pixel$^{-1}$). 

The absorber's host galaxy was expected to lie close to the quasar sightline given its strong absorption lines (W$_r$($\lambda$2796) = 2.43 $\pm$ 0.06 \AA), based on the known anti-correlation between Mg\II$ $ equivalent widths and impact parameters (\citealt{Chen2010, Nielsen2013a} and references therein). Figure \ref{fig:F140W} shows the F140W image where the scale is stretched to reveal a companion very close to the quasar sightline (projected distance $\sim$ 0.65$''$), being the best candidate for the host galaxy of the J1211+0833 absorber. Several objects are also present in the same field within $\sim$ 5$''$ ($\sim$42 kpc at $z$$_{\rm abs}$) of the quasar. The next steps are to remove contaminations and extract 1D spectra to derive the redshifts of the close companion and nearby objects.

\section{Contamination modeling, spectral extraction, and redshift confirmation}
\label{sec:reduction}

Due to the close proximity of the bright quasar, it is critical to model and subtract the quasar contamination. To model the 2D spectrum of the quasar or any given object in the direct image, we first compute the trace and dispersion parameters that define the mapping of a single pixel in the reference (direct) image into a one-pixel-wide spectrum in the dispersed image (i.e., the 2D grism image). The full 2D model is generated by shifting and adding this elemental spectrum scaled by the observed brightness in the direct image \citep{Brammer2012, Momcheva2016}. 

\begin{figure*}
\centering
{\includegraphics[width=11.5cm, height=10.9cm]{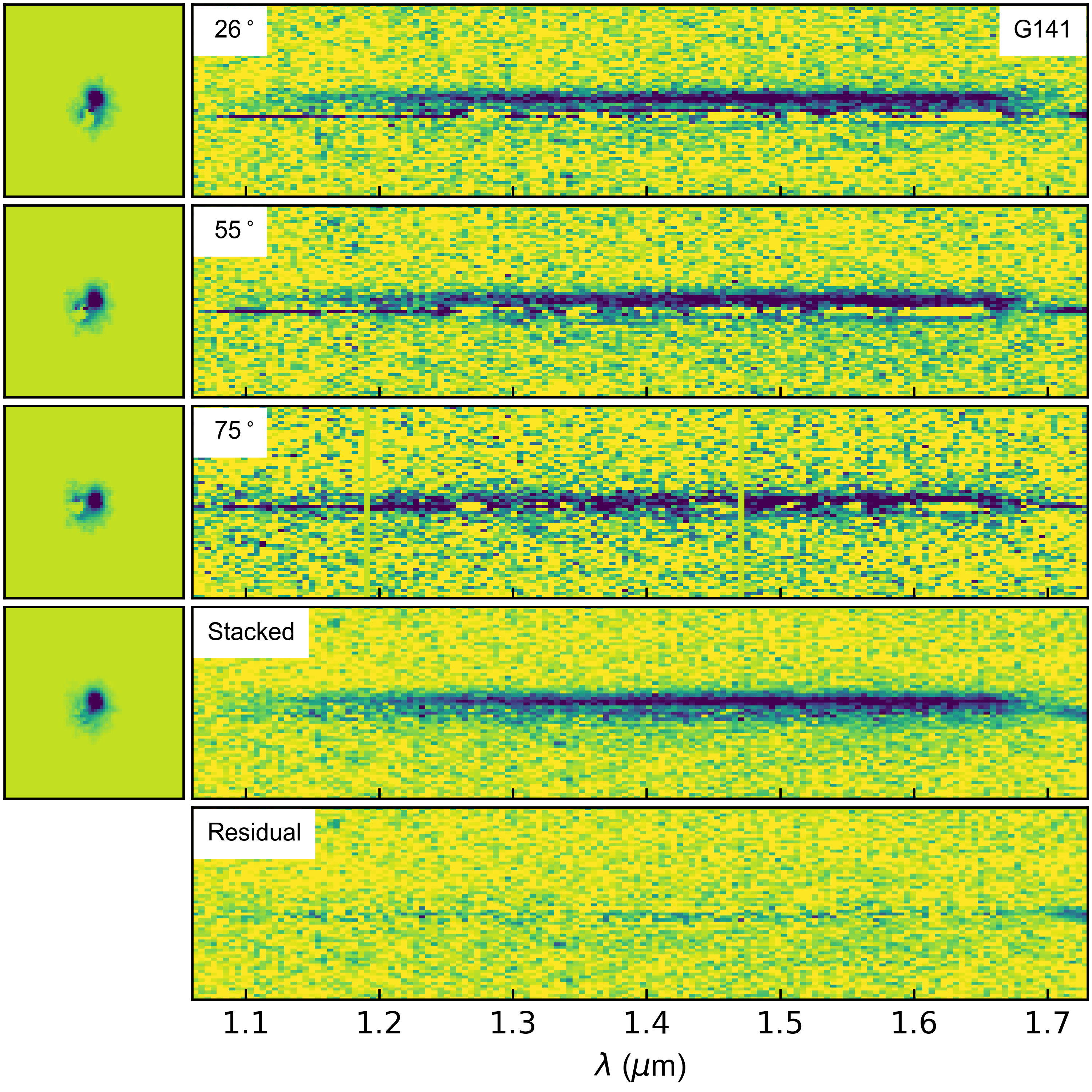}}   
{\includegraphics[width=7.1cm, height=6.97cm]{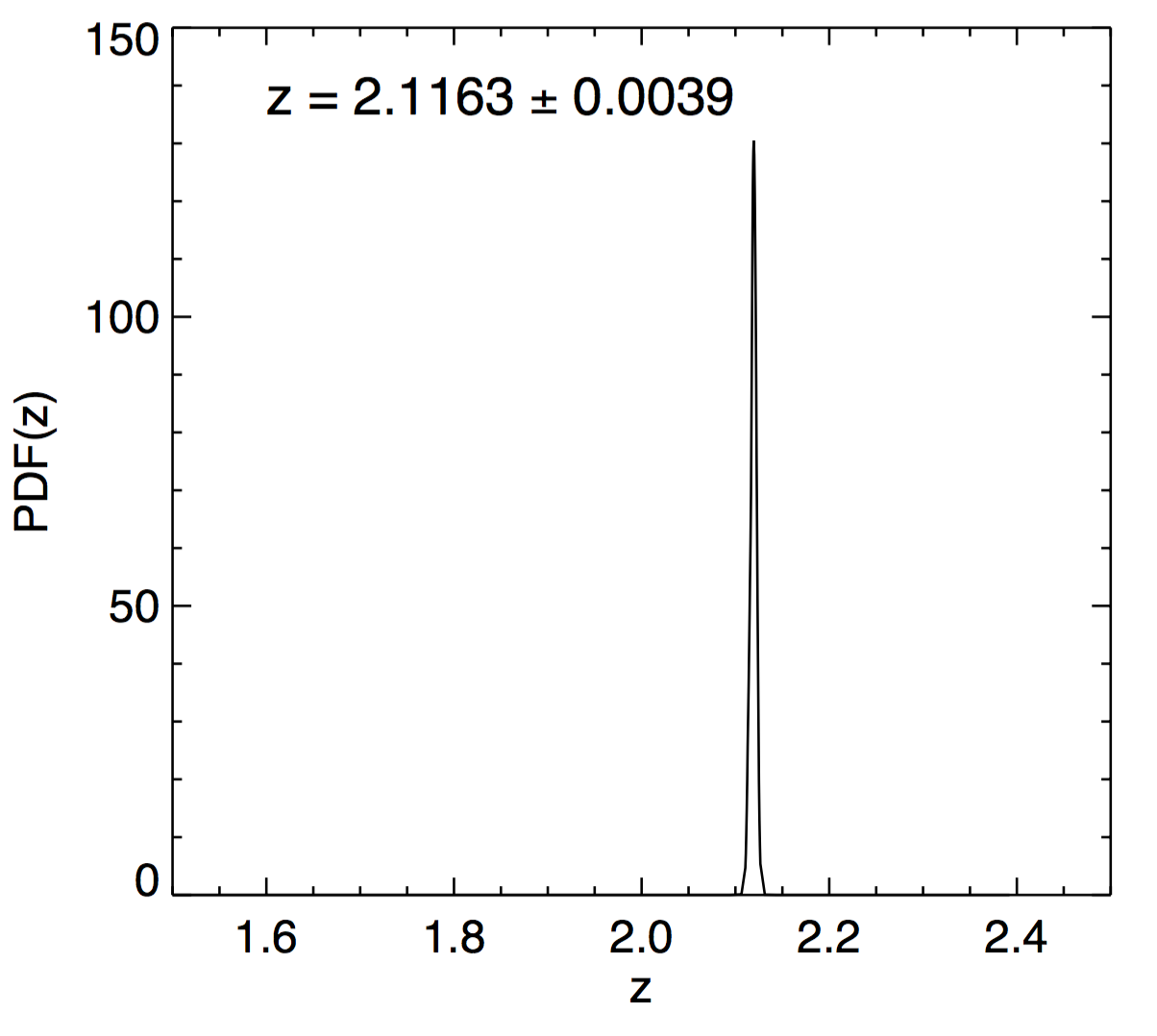}}  
{\includegraphics[width=10.2cm, height=7.1cm]{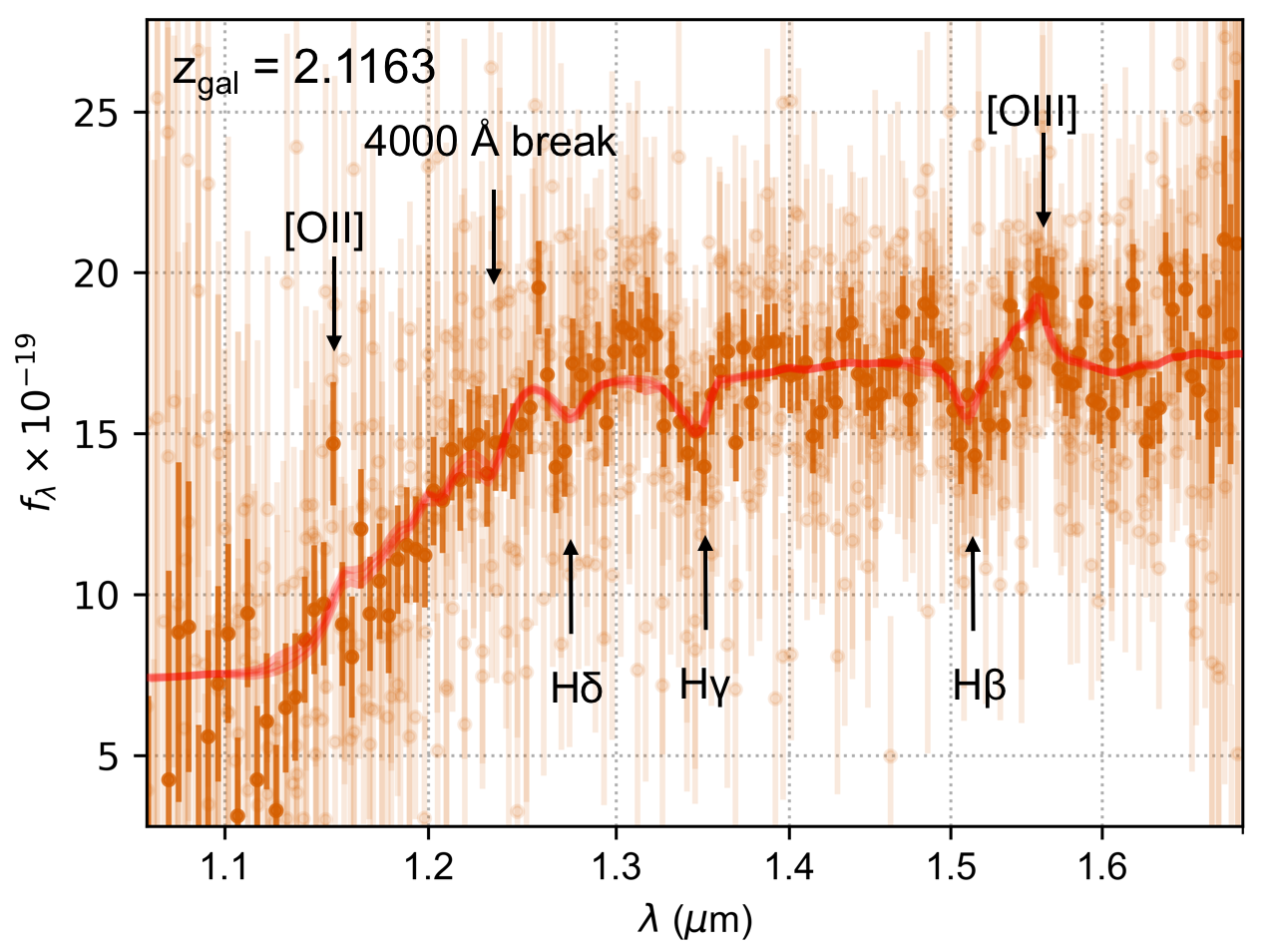}}    
\caption{{\it Top:} The first three rows show the residuals after subtracting the quasar in F140W (left) and G141 (right) in each visit. The dispersion position angles for the three visits are 26$^\circ$, 55$^\circ$, and 75$^\circ$ respectively. The fourth row shows the stacked image (left) and stacked 2D grism spectrum (right). The last row shows the overall residuals after subtracting the best-fit 2D spectrum for the host galaxy. {\it Bottom Left:} The redshift probability distribution. The best-fit redshift is at $z$ = 2.1163. {\it Bottom Right:} The extracted 1D spectrum for the 2DA host galaxy. The faint orange data points display the 1D spectra extracted from each of the six individual grism exposures, and the red curve is the best-fit template to the binned data (dark orange; binned to the native pixel scale of 46.5 \AA$ $ pixel$^{-1}$).}
\label{fig:data}
\end{figure*}

We use the effective point spread function (ePSF\footnote{http://www.stsci.edu/hst/wfc3/analysis/PSF}; \citealt{Anderson2016}) to model the quasar direct image and use the templates in the Grizli database to fit and subtract the 2D PSF trace iteratively. The top panel in Figure \ref{fig:data} displays the residuals after subtracting the quasar in each visit (first three rows). The quasar-subtracted direct and grism images from the three visits are re-drizzled for modeling the companion's 2D spectrum. The fourth row displays the stacked image and 2D spectrum from the three visits. The last row shows the reasonably clean residuals (the positive pixels are due to the residuals from the quasar subtraction) after subtracting the best-fit model for the companion. We determine the companion redshift by fitting the Flexible Stellar Population Synthesis templates (FSPS;\citealt{Conroy2009,Conroy2010}) to the quasar-subtracted exposure-level 2D spectra over a grid of redshifts. The bottom left panel of Figure \ref{fig:data} shows the result of this fit, indicating a prominent peak with maximum posterior probability of $z$ = 2.1163 $\pm$ 0.0039, which is in excellent agreement with the redshift of the 2DA ($\Delta$$z$ = $z$$_{gal}$ - $z$$_{abs}$ = 0.0003). This close companion is indeed responsible for the strong 2DA. The bottom right panel shows an optimal 1D extraction \citep{Horne1986} of the companion spectrum (orange points with uncertainties) and the best-fit FSPS model at the derived redshift (red curve).  The properties of the rest-frame optical spectrum of the absorber, which shows a continuum break and Balmer absorption lines, is discussed in more detail in Section \ref{section4.4}.

We further extract spectra and fit redshifts for the nearby objects in the same manner. G1 and G2 are at redshifts close to the companion's redshift ($z$$_{G1}$ = 2.1188 $\pm$ 0.0064, $z$$_{G2}$ = 2.2433 $\pm$ 0.0032). We have checked the deep VLT/UVES spectrum \citep{Ma2015} but no absorption features were found for these two galaxies.

\section{Discussion}
\label{sec:discussion}

We have demonstrated that the high-resolution {\it HST} imaging successfully recovers the absorber's host galaxy and enables extraction of the grism spectrum to confirm the redshift. The direct imaging and grism spectroscopy provide ample information about the properties of the host galaxy (Table \ref{tab}), including the impact parameter, morphology, SFR, SFR surface density, and stellar mass. These are the fundamental parameters that can be used to compare with other absorption-selected galaxies (e.g., DLAs, Mg\II$ $ absorbers) and place this population in context with the well-established galaxy populations in emission.

\begin{table*}
\centering
\caption{Properties of the J1211+0833 absorber's host galaxy}
\begin{tabular}{lc}
\hline
\hline
      &   {\it HST}/WFC3 F140W	 \\
\hline      
      Coordinate (RA; Dec)     & 12:11:43.44, +08:33:49.5 \\
      Magnitude (AB)    & 21.56 $\pm$ 0.10 \\
      S\'ersic index (n)   & 0.92 $\pm$ 0.10 \\
      Effective semi-major axis (R$_e$)& 0.27 $\pm$ 0.01$''$ or 2.24 $\pm$ 0.08 kpc  \\
      Circularized effective radius (R$_{e,c}$)& 1.91 $\pm$ 0.08 kpc\\
      Axis ratio (b/a) & 0.73 $\pm$ 0.03 \\
      Impact parameter (D) & 5.5 kpc\\
   \hline     
	      &  {\it HST}/WFC3 G141  \\
\hline	      
Galaxy redshift ($z$$_{gal}$)    & 2.1163 $\pm$ 0.0039 \\
$[OIII]$ line flux  & (6.2 $\pm$ 1.7)$\times$10$^{-17}$ erg s$^{-1}$cm$^{-2}$  \\
$[OII]$ line flux 3$\sigma$ upper limit & $<$ 7.9 $\times$ 10$^{-17}$ erg s$^{-1}$cm$^{-2}$ \\   
Unobscured SFR    (Chabrier IMF)                                     & 9.4 $\pm$ 2.6  M$_{\odot}$yr$^{-1}$ \\ 
$\Sigma$$_{\rm SFR}$     & 0.4 $\pm$ 0.1 M$_{\odot}$yr$^{-1}$kpc$^{-2}$\\
M$_*$ & (3 - 7) $\times$ 10$^{10}$ M$_{\odot}$ \\
\hline
\end{tabular}
\label{tab}
\end{table*}

\subsection{Morphology}

We perform surface brightness profile fitting on the F140W image using GALFIT \citep{Peng2010}. We derive the effective radius ($R_e$), S\'ersic index (n), and axis ratio (b/a). The measurements and uncertainties from GALFIT are presented in Table \ref{tab}. The S\'ersic index of 0.92 $\pm$ 0.10 is consistent with a disk morphology. The effective radius from GALFIT is along the semi-major axis, and we convert it to a circularized effective radius $R_{e,c}$ = $R_e$$\sqrt{b/a}$ = 1.91 $\pm$ 0.08 kpc using the measured axis ratio of 0.73 $\pm$ 0.03. These morphological parameters are in the range of the values of the $z$ $>$ 1 DLA and Mg\II$ $ galaxies in the literature \citep{Lundgren2012, Krogager2013}.

\subsection{Impact paramter}

\begin{figure*}
\centering
{\includegraphics[width=8.95cm, height=6.5cm]{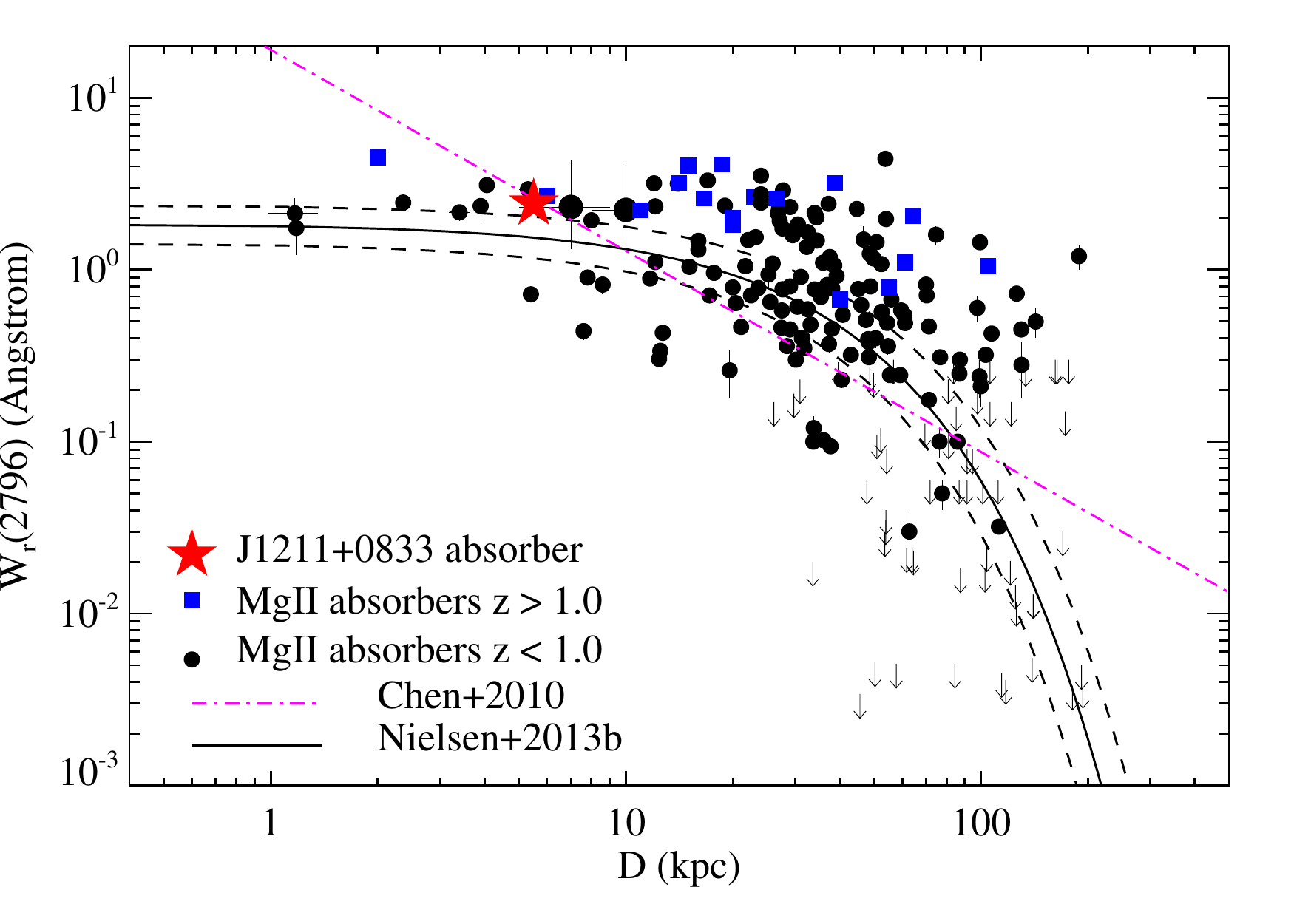}}  
{\includegraphics[width=8.95cm, height=6.5cm]{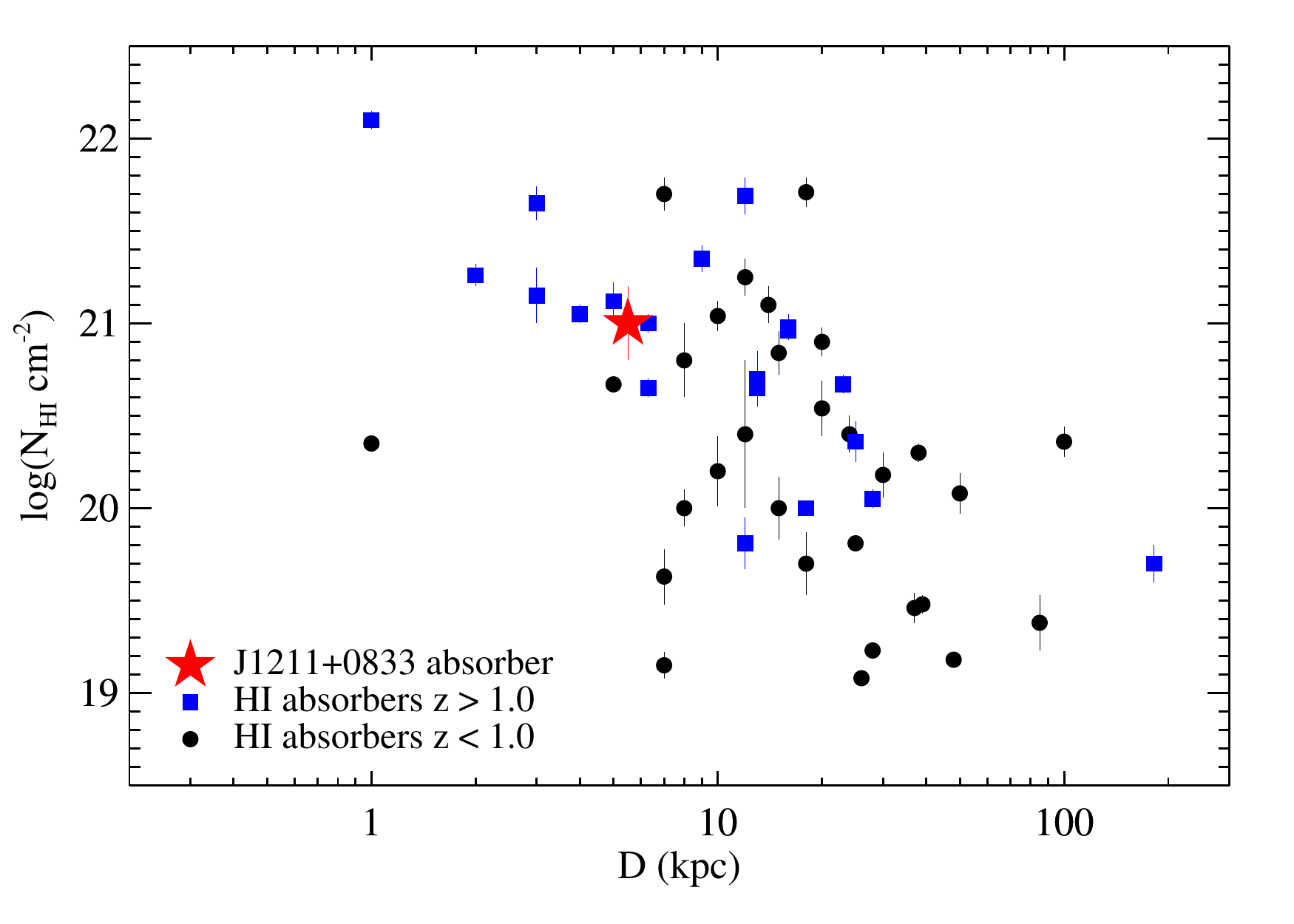}} 
\caption{{\it Left:} Observed anti-correlation between the rest-frame equivalent width of Mg\II$ $ absorption, W$_r$($\lambda$2796), and galaxy impact paramter, D. The individual measurements from the literature include \cite{Bouche2007, Bouche2012}, \cite{Lundgren2012}, \cite{Nielsen2013a}, \cite{Kacprzak2013}, and \cite{Joshi2017}. The pink dash-dotted line is the power-law fit obtained by \cite{Chen2010}. The solid curve is a log-linear maximum likelihood fit given by \cite{Nielsen2013b} with the associated 1$\sigma$ denoted by the dashed curves.  {\it Right:} Anti-correlation between logN(H\I) and impact parameter. The black circles and blue squares are the detected DLA and subDLA host galaxies in the literature at $z$ $<$ 1 and $z$ $>$ 1 respectively, compiled by \cite{Rahmani2016}.}
\label{fig:pho}
\end{figure*}

\subsubsection{W$_r$($\lambda$2796) and impact parameter}

The anti-correlation between  the rest-frame equivalent width of Mg\II$ $ absorption and the impact parameter has been well-established (e.g., \citealt{Bergeron1991,Steidel1995,Chen2010,Rao2011,Nielsen2013a,Nielsen2013b}), albeit with a large scatter. \cite{Chen2010} describes the relation using a simple power-law, while \cite{Nielsen2013a,Nielsen2013b} have shown a 7.9$\sigma$ anti-correlation well represented by a log-linear relation. Figure \ref{fig:pho} shows the individual measurements at 0.10 $<$ $z$ $<$ 2.35 in the literature and the best-fit relations by \cite{Chen2010} and \cite{Nielsen2013a,Nielsen2013b}. 

The host galaxy of the J1211+0833 2DA is located at an impact parameter of 5.5 kpc, which is the smallest angular separation among the Mg\II$ $ absorbers at $z$ $>$ 2 directly detected and spectroscopically confirmed in emission. This absorber presents ultra-strong Mg\II$ $ absorption lines (W$_r$($\lambda$2796) = 2.43 $\pm$ 0.06 \AA; \citealt{Ma2015}) in close agreement with the power-law relation obtained by \cite{Chen2010}.

\subsubsection{N(H\I) and impact parameter}

The right panel in Figure \ref{fig:pho} displays N(H\I) as a function of impact parameter of DLA and subDLAs for the spectroscopically confirmed absorbing galaxies compiled by \cite{Rahmani2016}. The DLAs/subDLAs at $z$ $>$ 1 on average show higher N(H\I) and smaller impact parameters than those at $z$ $<$ 1. At impact parameters higher than 20 kpc, there are no DLAs with logN(H\I/cm$^{-2}$) $>$ 21.0. The host galaxy of the 2DA is located among the ones with high N(H\I) and small impact parameters. A few high-$z$ DLA galaxies are detected at very small impact parameters ($<$ 5 kpc; \citealt{Moller1993,Krogager2012,Noterdaeme2012}).

\subsection{SFR and SFR surface density}

\begin{figure}
\centering
{\includegraphics[width=9.0cm, height=6.5cm]{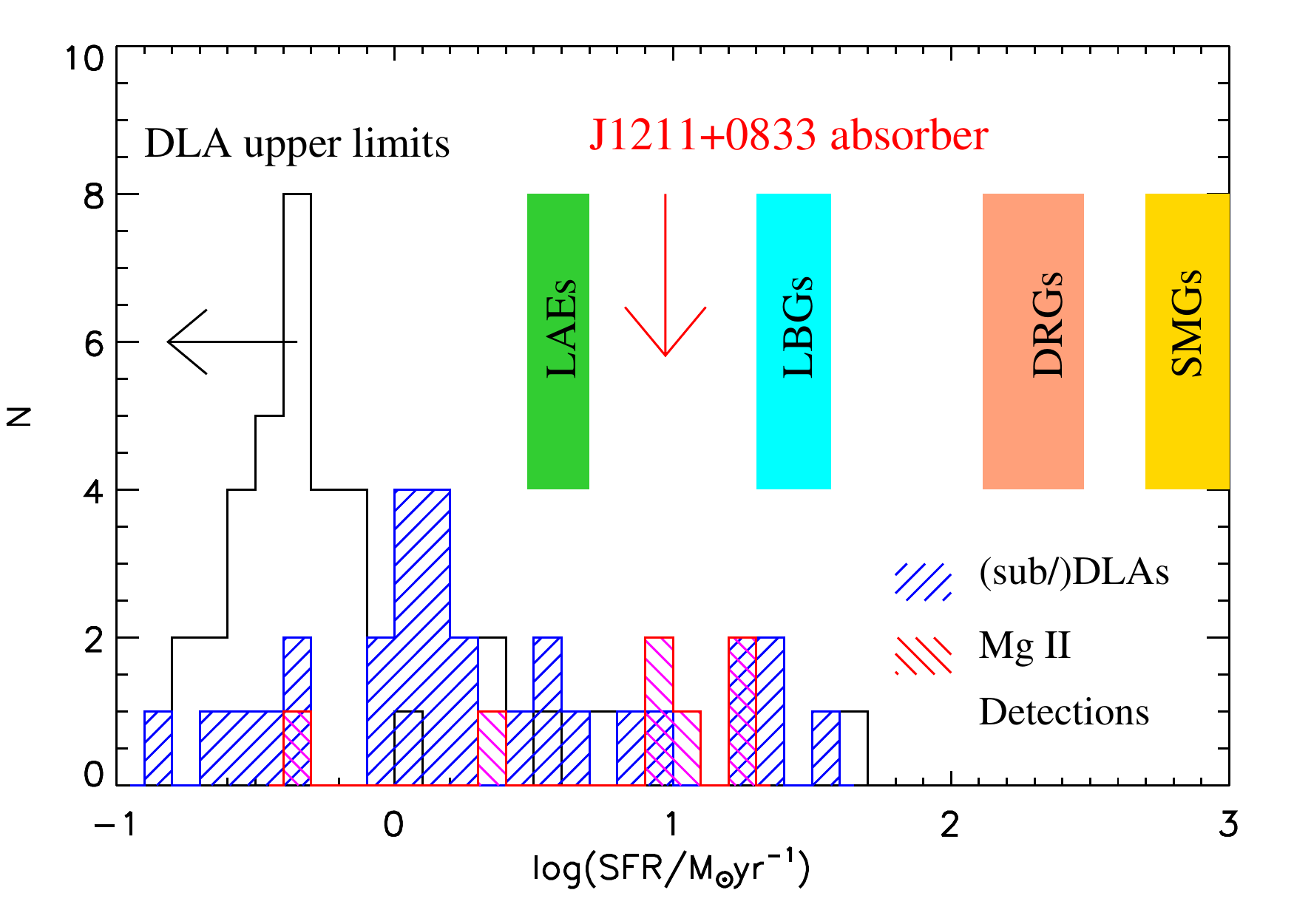}}    
\caption{SFRs of Mg\II$ $, H\I$ $ galaxies, and emission selected galaxies at $z$ $\sim$ 2-3. The blue, red, and black histograms show the SFRs of DLA/subDLA galaxies, Mg\II$ $ galaxies, and DLA non-detections \citep{Bouche2012, Lundgren2012, Fumagalli2015, Rahmani2016}. The color bars represent the median SFRs of LAEs, LBGs, DRGs, and SMGs at similar redshifts \citep{Reddy2006, Reddy2010, Hagen2016}. The SFR of the J1211+0833 2DA host galaxy is between those of LAEs and LBGs.}
\label{fig:sfr}
\end{figure}

The overall shape of the extracted grism spectrum is dominated by a continuum with a prominent Balmer/4000 \AA$ $ break and weak emission lines, [O\III]$\lambda$5007 and [O\II]$\lambda$3727. We measure the line fluxes to be (6.2 $\pm$ 1.7)$\times$10$^{-17}$ erg s$^{-1}$cm$^{-2}$ for [O\III] and $<$ 7.9 $\times$10$^{-17}$ erg s$^{-1}$cm$^{-2}$ (3$\sigma$ upper limit) for [O\II]. Emission line fluxes are determined from the normalization of the emission line templates, and their uncertainties are derived from the covariance matrix of the 2D template models weighted by the pixel variances.

[O\II] and [O\III] have been used as SFR indicators whenever H$\alpha$ is not available, albeit with uncertainties as large as a factor of 2-4. We convert the [O\III] luminosity (assuming [O\III]/H$\alpha$ = 1) to SFR using the H$\alpha$-based SFR calibration of \cite{Kennicutt1998} converted to a \cite{Chabrier2003} initial mass function (IMF), SFR  (M$_{\odot}$yr$^{-1}$) = 4.4 $\times$ 10$^{-42}$ L$_{H\alpha}$/erg s$^{-1}$ = 9.4 $\pm$ 2.6 M$_{\odot}$yr$^{-1}$. The [O\II]-based SFR is calculated as SFR  (M$_{\odot}$yr$^{-1}$) = 7.8 $\times$ 10$^{-42}$ L$_{[OII]}$/erg $s^{-1}$ $<$ 21 M$_{\odot}$yr$^{-1}$ (3$\sigma$ upper limit). This SFR is in the range of the SFRs of Mg\II$ $ host galaxies at $z$ $\sim$ 2 with W$_r$($\lambda$2796) $>$ 2 \AA$ $ \citep{Bouche2012, Lundgren2012}. We compare the SFRs of Mg\II$ $ and H\I$ $ galaxies with emission-selected galaxies at $z$ $\sim$ 2-3 in Figure \ref{fig:sfr}. The J1211+0833 2DA's host galaxy has a SFR significantly higher than most of the DLA/subDLA galaxies and lies between the median values of LAEs and LBGs. 

We further calculate the SFR surface density given the size measurement from the F140W imaging as

\begin{equation}
\Sigma SFR   (M_{\odot}yr^{-1} kpc^{-2}) = SFR/(2\pi R^2_{e,c})
\end{equation}
where R$_{e,c}$ is the circularized effective radius of the host galaxy. The $\Sigma$SFR is within the range of the Mg\II$ $ host galaxies in \cite{Lundgren2012}.

\subsection{Dn(4000), age, stellar mass, Virial radius, and star formation history}
\label{section4.4}

The break at 4000 \AA$ $ arises because of the accumulation of a large number of metal lines in a narrow wavelength region. The strength of the 4000 \AA$ $ break is measured as the flux ratio between the wavelength windows adjacent to the 4000 \AA$ $ break. We use the Dn(4000) index defined as the ratio between 3850-3950 \AA$ $ and 4000 - 4100 \AA$ $ \citep{Balogh1999}. The Dn(4000) index increases as a function of stellar population age and provide a characteristic age for the galaxy \citep{Kauffmann2003a}. We measure a Dn(4000) index for the 2DA host galaxy of 1.3 $\pm$ 0.3, indicating the stellar population is about 1 Gyr old. We estimate the stellar mass to be in the range of log(M$_*$/M$_{\odot}$)= 10.5 -10.8 based on two correlations: the Dn(4000)-stellar mass relation \citep{Kauffmann2003b,Hernan2013} and the mass-size relation \citep{Krogager2014}. This stellar mass is within the range predicted by the stellar mass estimators based on absorption properties (Equations 5 \& 6 in \citealt{Ma2018}). Since we only have one photometric data point at F140W and a narrow window of spectral coverage, multi-wavelength spectral energy distribution modeling with more photometric and/or spectral coverage would improve the estimates. We further convert the stellar mass to halo mass \citep{Behroozi2013} to derive the Virial radius of the host galaxy halo to be $\sim$ 80 - 100 kpc. The background quasar is already probing the ISM of the 2DA host galaxy while most quasar absorption line systems trace the circumgalactic or intergalactic medium. 

The [O\III] and [O\II] emission lines, although weak, suggest ongoing star formation in the host galaxy. On the other hand, moderate Balmer absorption lines are also present in the spectrum, which are signs of increased fraction of A or F type stars in a recent starburst. The type of galaxy hosting the 2DA emerges as being in the transition from a star-forming galaxy to a quiescent galaxy where star formation has been quenched. Galaxies caught in the act of transition are either in the final stage of a starburst or have recently quenched star formation, and the so-called post-starburst galaxies are one such class \citep{Wild2016}. The J1211+0833 absorber's host galaxy is likely in the early stage of transition, i.e., the final starburst phase.  \newline

A coherent picture of the 2DA host galaxy has been pieced together with the properties derived from the absorption analysis \citep{Ma2015,Ma2017,Ma2018} and the emission counterpart studied in this work:  The host galaxy is indeed a chemically-enriched, evolved, massive, and star-forming disk-like galaxy, although it is likely in the final stage of star formation and is transitioning to a quiescent galaxy. The SFR and $M_*$ of the host galaxy are comparable to though slightly lower than those of typical galaxies at $z$ $\sim$ 2, i.e. LBGs, BzK, UV-selected galaxies, and distant red galaxies (DRGs) \citep{Reddy2006, Reddy2010, Buat2012}. Detecting the host galaxy at multi-wavelengths would enable a fair comparison with different emission-selected galaxy populations. It is also promising that far-infrared emission lines such as CO and C\I$ $ and dust continuum emission can be detected with ALMA given their presence in absorption \citep{Ma2015}. It would be of great interest to compare its gas and dust properties to those of mid-IR and FIR/submmillimeter selected galaxies, especially the polycyclic aromatic hydrocarbon (PAH) bands at mid-IR, which are considered as a promising candidate for the 2175 \AA$ $ bump carriers (e.g., \citealt{Draine2003, Xiang2011}).

It is worth noting that the similar redshifts of the nearby galaxies indicate a group environment. The 2175 \AA$ $ bump is ubiquitous in our Milky Way Galaxy, which also resides in a group environment with multiple satellite galaxies. It is still too early to draw conclusions on whether this is a coincidence or the group environment does play a role in the formation and presence of the 2175 \AA$ $ bump carriers.

\section{Summary and conclusions}
\label{sec:conclusion}

We have detected the emission counterpart or the host galaxy of a 2DA for the first time using WFC3 IR direct imaging and grism spectroscopy. We have subtracted the quasar contamination and performed template fitting on the extracted spectrum to confirm the redshift. The spectroscopically confirmed host galaxy is located at a small impact parameter of 5.5 kpc. The morphological parameters derived from the F140W image indicate that it is a disk-like galaxy with an axis ratio of 0.73. The spectral analysis (weak [O\III] and [O\II] emission lines, moderate 4000 \AA$ $ break and Balmer absorption lines) shows that the host galaxy contains an evolved stellar population with small amount of current star formation. The host galaxy is found to be a chemically-enriched, evolved, massive, and star-forming disk-like galaxy that is in the transition from a blue star-forming galaxy to a red quiescent galaxy. More detections like this in the future are required to establish this population as a whole in the context of galaxy formation and evolution.

\section*{acknowledgments}

We thank the anonymous referee for the constructive comments that have improved the manuscript. J.M. is grateful to Anna Nierenburg for local technical support. This work is based on observations made with the NASA/ESA {\it Hubble Space Telescope} obtained at the Space Telescope Science Institute, which is operated by the Association of Universities for Research in Astronomy, Inc., under NASA contract NAS 5-26555. We acknowledge support from the {\it HST} GO-14200 grant. This research has made use of NASA's Astrophysics Data System.

\end{document}